\documentclass[journal]{IEEEtran}

\usepackage{amssymb,epic,eepic}

\usepackage[dvips]{color}
\usepackage{graphicx}

\usepackage{pst-plot, pstricks}

\usepackage[arrow, matrix, curve]{xy}

\begin{document}
\title{Quantum Bit Strings and Prefix-Free Hilbert Spaces}

\newcommand{\Tr}{{\rm Tr}}
\newcommand{\St}{{\mathcal{S}}}
\newcommand{\hr}{{\cal H}}
\newcommand{\cnq}{{\hr_n^\mathbb{Q}}}
\newcommand{\cn}{{\hr_n}}
\newcommand{\C}{{\mathbb C}}
\newcommand{\R}{{\mathbb R}}
\newcommand{\N}{{\mathbb N}}
\newcommand{\idn}{\mathbf{1}}
\newcommand{\Z}{{\mathbb Z}}
\newcommand{\x}{{\mathbf x}}
\newcommand{\eps}{{\varepsilon}}
\newcommand{\om}{\omega}
\newcommand{\kk}{{\mathbf k}}
\newcommand{\n}{{\mathbf n}}
\newcommand{\y}{{\mathbf y}}
\newcommand{\s}{{\{0,1\}^*}}
\newtheorem{theorem}{Theorem}[section]
\newtheorem{lemma}[theorem]{Lemma}
\newtheorem{corollary}[theorem]{Corollary}
\newtheorem{definition}[theorem]{Definition}
\newtheorem{proposition}[theorem]{Proposition}
\newtheorem{conjecture}[theorem]{Conjecture}
\newtheorem{example}[theorem]{Example}
\newcommand{\nix}{{\rule{0pt}{2pt}}}
\newcommand{\qedd}{{\nix\nolinebreak\hfill\hfill\nolinebreak$\Box$}}
\newcommand{\qed}{{\qedd\par\medskip\noindent}}
\newcommand{\lineclear}{{\rule{0pt}{0pt}\nopagebreak\par\nopagebreak\noindent}}

\author{Markus M\"uller and Caroline Rogers%
\thanks{M. M\"uller is with the Max Planck Institute for Mathematics in the Sciences,
Inselstr. 22, D-04103 Leipzig, Germany (e-mail: mueller@math.tu-berlin.de).}
\thanks{C. Rogers is with the Department of Computer Science,
University of Warwick, Gibbett Hill Road, Coventry, CV4 7AL, United Kingdom (e-mail: caroline@dcs.warwick.ac.uk).
}%
}%
\markboth{Quantum Bit Strings and Prefix-free Hilbert Spaces}{}

\maketitle

\begin{abstract}
We give a mathematical framework for manipulating indeterminate-length quantum bit strings.
In particular, we define prefixes, fragments, tensor products and concatenation of such strings of qubits,
and study their properties and relationships.

The results are then used to define prefix-free Hilbert spaces
in a more general way than in previous work, without assuming the existence of a basis of length eigenstates.
We prove a quantum analogue of the Kraft inequality, illustrate the results with some examples and
discuss the relevance of prefix-free Hilbert spaces for lossless compression.
\end{abstract}


\IEEEpeerreviewmaketitle

\section{Introduction and Formalism}
\PARstart{I}n classical information theory, prefix codes play a crucial role in coding theory
and compression \cite{CoverThomas}, as well as in algorithmic information theory \cite{LiVitanyi}.
Thus, it is a natural question how the concept of a prefix code can be generalized
to quantum information theory.

Quantum prefix codes have first been defined by Schumacher and Westmoreland~\cite{SchumacherWestmoreland}.
They also proved a quantum version of the Kraft inequality, and showed that quantum prefix codes
can be used for lossless quantum compression. A basic problem for lossless compression in the quantum case is
that different code words may have different lengths. Say, if one message has code word $00$, and another one has
code word $1111$, then we may also have a superposition of both messages, leading to code words like
\[
   |\psi\rangle:=\frac 1 {\sqrt{2}}\left(|00\rangle-|1111\rangle\right).
\]
Thus, lossless compression naturally leads to quantum bit strings (qubit strings) which are superpositions
of classical bit strings of different lengths. There is a lot of previous work on such ``indeterminate-length
qubit strings'' and the related problem of lossless quantum compression,
see for example \cite{BostroemFelbinger}, \cite{AhlswedeCai}, \cite{Braunstein}, and
\cite{Koashi}.

Formally, qubit strings like $|\psi\rangle$ can be defined as elements of a
Hilbert space $\hr_\s$, the string space. It contains the classical binary strings
$\s=\{\lambda,0,1,00,01,\ldots\}$ as orthonormal basis vectors ($\lambda$ denotes the empty string).
Starting with $\C^2$, the usual $n$-qubit space is the $n$-fold tensor product
$\left(\C^2\right)^{\otimes n}$. Allowing every possible length $n\in\N_0$, we define the string space $\hr_\s$ as the linear
span of all the $n$-qubit spaces, that is, as the direct sum
\[
   \hr_\s:=\bigoplus_{n=0}^\infty \left(\C^2\right)^{\otimes n}.
\]
Instead of working directly on $\hr_\s$, Schumacher and Westmoreland~\cite{SchumacherWestmoreland}
used a construction called zero-extended form: indeterminate-length
qubit strings like $|\psi\rangle$ are filled up with zeroes, until they have a fixed length $n$, i.e. they are ordinary
vectors in $\left(\C^2\right)^{\otimes n}$. The space of allowed
code words is then spanned by self-delimiting quantum strings $|\varphi_i\rangle$ (with zeroes appended).
Each basis code word $|\varphi_i\rangle$
contains in its superposition only classical strings of some fixed length $\ell_i$
(the qubit strings $|\varphi_i\rangle$ are ``length eigenstates'').

Consequently, in that approach, every quantum prefix code has a basis of length eigenstates.
However, this is some kind of artificial restriction, resulting from the construction of zero-extended forms.
It is an interesting question if this restriction can be lifted, and what can be gained
by a potential generalization.

In this paper, we give such a generalization by defining quantum prefix codes
directly on the string space $\hr_\s$. This also has the advantage to refer to indeterminate-length qubit strings
like $|\psi\rangle$ in a more natural and direct way, without zero-extended forms.
Meanwhile, we define and analyze how such qubit strings can be manipulated:
we define prefixes, concatenations, restrictions and tensor products.

The paper consists of two parts: Section~\ref{SecFormal} is rather abstract; it contains the formal
preparations and definitions. In contrast, Section~\ref{SecIntuitive} is more easy and intuitive:
it studies the properties of prefix-free sets on the string space. In particular, it contains a quantum
generalization of the Kraft inequality for arbitrary prefix-free Hilbert spaces.

We conclude this introduction by explaining some more formalism.
We denote the length of a classical string $s\in\s$ by $\ell(s)$.
The subspace of $\hr_\s$ which is spanned by the classical strings of length less or equal than $k$
is denoted $\hr_{\leq k}$, i.e. $\hr_{\leq k}:=\bigoplus_{n=0}^k \left(\C^2\right)^{\otimes n}$. For example,
$|\psi\rangle\in\hr_{\leq 4}$, but $|\psi\rangle\not\in\hr_{\leq 3}$.
The space $\hr_\s$ contains a natural ``length observable'' $\Lambda$, an unbounded operator, which  is defined by
linear extension (with maximal domain of definition) of
\[
   \Lambda|s\rangle:=\ell(s)|s\rangle\mbox{ for all }s\in\s.
\]
As we will see later, prefixes of qubit strings can be mixed quantum states. For this reason,
not only state vectors $|\varphi\rangle\in\hr_\s$, but also density operators $\rho$ on $\hr_\s$
will be called qubit strings. We define the ``base length'' $\ell(\rho)$ of some qubit string $\rho$ as
\[
   \ell(\rho):=\max\{\ell(s)\,\,|\,\,s\in\s, \enspace\langle s|\rho|s\rangle>0\}.
\]
A qubit string $\rho$ is called ``length eigenstate'' if $\langle s |\rho|s\rangle> 0$ only for strings $s\in\s$
with fixed length $\ell(s)=\ell(\rho)$.
Moreover, we define the ``average length'' $\bar\ell(\rho)$ as
\[
   \bar\ell(\rho):=\Tr\left(\rho\Lambda\right).
\]
If $U\subset\hr$ is some closed subspace of a Hilbert space $\hr$, then $\mathbb{P}(U)$ denotes
the orthogonal projector onto $U$. The cardinality of some set $I$ is denoted $|I|$. By $\idn$,
we denote the identity map, and $\idn_\hr$ is the identity map on some Hilbert space $\hr$.
The symbol $\circ$ is for concatenation, e.g. $10\circ 01=1001$. The set of bounded operators on a Hilbert
space $\hr$ is denoted $\mathcal{B}(\hr)$, and the set of density operators on $\hr$ is $\mathcal{S}(\hr)$.

\section{Prefixes, Fragments, Tensor Products, and Concatenation of Quantum Bit Strings}
\label{SecFormal}
Given some qubit string $\varphi\in\hr_\s$, then how can we define its prefix $\varphi_1^n$,
i.e. the fragment of the first $n$ qubits of $\varphi$?
For some quantum state on a tensor product of Hilbert spaces, the restriction of that state
to a subsystem is given by the partial trace, and so should be $\varphi_1^n$. But partial
traces are only defined if there is some tensor product structure; apriori, $\hr_\s$ does not
have any tensor product structure.

We can solve this problem by recalling that we would like to use indeterminate-length qubit strings for
quantum computation. For this reason, we must somehow embed such qubit strings in the memory of a quantum computer.
This idea led Schumacher and Westmoreland to the definition of zero-extended forms \cite{SchumacherWestmoreland}.

Instead of appending zeroes,
we can recall how classical bit strings are written on the tape of a Turing machine (TM) \cite{LiVitanyi}.
The tape of a TM consists of infinitely many cells, indexed by integers. Each cell can either carry a $0$, a $1$, or a blank symbol $\#$.
If a bit string $s$ of length $\ell$ is input into the TM, then the first $\ell$ cells are filled with the bits of $s$,
and all the other cells are blank. In every step of the computation, all but a finite number of cells
carry the blank symbol $\#$.

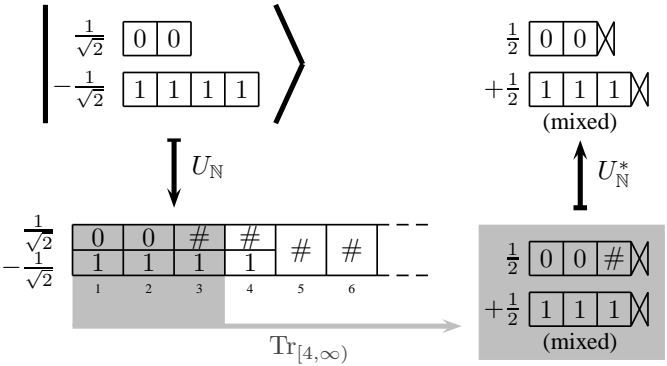
\begin{figure}[!hbt]
\psset{unit=0.45cm}
\begin{center}
\begin{pspicture}(-1,-2.5)(26,8)
   \psline[linewidth=3,linecolor=lightgray](1,0)(5.5,0)
   \psline[linewidth=4,linecolor=lightgray](13,-0.5)(18.5,-0.5)
   
   {\psset{linewidth=0.15}\psline(0.2,8)(0.2,4.5)}
   \rput(1.5,7){{$\frac 1 {\sqrt{2}}$}}
   \rput(1.2,5.5){{$-\frac 1 {\sqrt{2}}$}}
   \rput(3,7){{$0$}}
   \rput(4,7){{$0$}}
   {\psset{linewidth=0.05}\psline(2.5,7.5)(4.5,7.5)}
   {\psset{linewidth=0.05}\psline(2.5,6.5)(4.5,6.5)}
   {\psset{linewidth=0.05}\psline(2.5,6.5)(2.5,7.5)}
   {\psset{linewidth=0.05}\psline(3.5,6.5)(3.5,7.5)}
   {\psset{linewidth=0.05}\psline(4.5,6.5)(4.5,7.5)}
   \rput(3,5.5){{$1$}}
   \rput(4,5.5){{$1$}}
   \rput(5,5.5){{$1$}}
   \rput(6,5.5){{$1$}}
   {\psset{linewidth=0.05}\psline(2.5,6)(6.5,6)}
   {\psset{linewidth=0.05}\psline(2.5,5)(6.5,5)}
   {\psset{linewidth=0.05}\psline(2.5,5)(2.5,6)}
   {\psset{linewidth=0.05}\psline(3.5,5)(3.5,6)}
   {\psset{linewidth=0.05}\psline(4.5,5)(4.5,6)}
   {\psset{linewidth=0.05}\psline(5.5,5)(5.5,6)}
   {\psset{linewidth=0.05}\psline(6.5,5)(6.5,6)}
   {\psset{linewidth=0.15}\psline(7,8)(7.8,6.25)}
   {\psset{linewidth=0.15}\psline(7.8,6.25)(7,4.5)}
   {\psset{linewidth=0.15}\psline{->}(4,4)(4,2)}
   {\psset{linewidth=0.1}\psline(3.8,4)(4.2,4)}
   \rput(5,3.2){{$U_\N$}}
   {\psset{linewidth=0.05}\psline(1,1.5)(10,1.5)}
   {\psset{linewidth=0.05}\psline(1,0)(10,0)}
   {\psset{linewidth=0.05}\psline(1,1.5)(1,0)}
   {\psset{linewidth=0.05}\psline(2.5,1.5)(2.5,0)}
   {\psset{linewidth=0.05}\psline(4,1.5)(4,0)}
   {\psset{linewidth=0.05}\psline(5.5,1.5)(5.5,0)}
   {\psset{linewidth=0.05}\psline(7,1.5)(7,0)}
   {\psset{linewidth=0.05}\psline(8.5,1.5)(8.5,0)}
   {\psset{linewidth=0.05}\psline(10,1.5)(10,0)}
   \rput(7.75,0.75){{$\#$}}
   \rput(9.25,0.75){{$\#$}}
   {\psset{linewidth=0.05}\psline(1,0.75)(7,0.75)}
   \rput(1.75,1.125){{$0$}}
   \rput(3.25,1.125){{$0$}}
   \rput(4.75,1.125){{$\#$}}
   \rput(6.25,1.125){{$\#$}}
   \rput(1.75,0.375){{$1$}}
   \rput(3.25,0.375){{$1$}}
   \rput(4.75,0.375){{$1$}}
   \rput(6.25,0.375){{$1$}}
   \rput(1.75,-0.4){{\tiny{1}}}
   \rput(3.25,-0.4){{\tiny{2}}}
   \rput(4.75,-0.4){{\tiny{3}}}
   \rput(6.25,-0.4){{\tiny{4}}}
   \rput(7.75,-0.4){{\tiny{5}}}
   \rput(9.25,-0.4){{\tiny{6}}}
   \rput(0,1.25){{$\frac 1 {\sqrt{2}}$}}
   \rput(-0.3,0.15){{$-\frac 1 {\sqrt{2}}$}}
   {\psset{linewidth=0.05,linestyle=dashed}\psline(10,1.5)(11.5,1.5)}
   {\psset{linewidth=0.05,linestyle=dashed}\psline(10,0)(11.5,0)}
   \psline[linewidth=0.15,linecolor=lightgray]{->}(1,-1.5)(12.5,-1.5)
   \rput(8,-2.2){{${\color{darkgray}\Tr_{[4,\infty)}}$}}
   
   \rput(14,0.5){{$\frac 1 2$}}
   {\psset{linewidth=0.05}\psline(14.5,1)(14.5,0)}
   \rput(15,0.5){{$0$}}
   \rput(16,0.5){{$0$}}
   \rput(17,0.5){{$\#$}}
   {\psset{linewidth=0.05}\psline(14.5,1)(17.5,1)}
   {\psset{linewidth=0.05}\psline(14.5,0)(17.5,0)}
   {\psset{linewidth=0.05}\psline(17.5,0)(17.5,1)}
   {\psset{linewidth=0.05}\psline(15.5,0)(15.5,1)}
   {\psset{linewidth=0.05}\psline(16.5,0)(16.5,1)}
   
   {\psset{linewidth=0.05}\psline(17.5,1)(18,0)}
   {\psset{linewidth=0.05}\psline(18,1)(17.5,0)}
   {\psset{linewidth=0.05}\psline(18,1)(18,0)}
   
   \rput(13.7,-1){{$+\frac 1 2$}}
   {\psset{linewidth=0.05}\psline(14.5,-0.5)(14.5,-1.5)}
   \rput(15,-1){{$1$}}
   \rput(16,-1){{$1$}}
   \rput(17,-1){{$1$}}
   {\psset{linewidth=0.05}\psline(14.5,-0.5)(17.5,-0.5)}
   {\psset{linewidth=0.05}\psline(14.5,-1.5)(17.5,-1.5)}
   {\psset{linewidth=0.05}\psline(17.5,-1.5)(17.5,-0.5)}   
   {\psset{linewidth=0.05}\psline(15.5,-1.5)(15.5,-0.5)}
   {\psset{linewidth=0.05}\psline(16.5,-1.5)(16.5,-0.5)}
   
   {\psset{linewidth=0.05}\psline(17.5,-0.5)(18,-1.5)}
   {\psset{linewidth=0.05}\psline(18,-0.5)(17.5,-1.5)}
   {\psset{linewidth=0.05}\psline(18,-0.5)(18,-1.5)}

   \rput(16,-2){\small(mixed)}
   
   \rput(17,3){{$U_\N^*$}}
   
   {\psset{linewidth=0.15}\psline{->}(16,2)(16,4)}
   {\psset{linewidth=0.15}\psline(15.8,2)(16.2,2)}

   \rput(14,7){{$\frac 1 2$}}
   {\psset{linewidth=0.05}\psline(14.5,7.5)(14.5,6.5)}
   \rput(15,7){{$0$}}
   \rput(16,7){{$0$}}
   {\psset{linewidth=0.05}\psline(14.5,7.5)(16.5,7.5)}
   {\psset{linewidth=0.05}\psline(14.5,6.5)(16.5,6.5)}
   {\psset{linewidth=0.05}\psline(15.5,6.5)(15.5,7.5)}
   {\psset{linewidth=0.05}\psline(16.5,6.5)(16.5,7.5)}
   
   {\psset{linewidth=0.05}\psline(16.5,7.5)(17,6.5)}
   {\psset{linewidth=0.05}\psline(17,7.5)(16.5,6.5)}
   {\psset{linewidth=0.05}\psline(17,7.5)(17,6.5)}
   
   \rput(13.7,5.5){{$+\frac 1 2$}}
   {\psset{linewidth=0.05}\psline(14.5,6)(14.5,5)}
   \rput(15,5.5){{$1$}}
   \rput(16,5.5){{$1$}}
   \rput(17,5.5){{$1$}}
   {\psset{linewidth=0.05}\psline(14.5,6)(17.5,6)}
   {\psset{linewidth=0.05}\psline(14.5,5)(17.5,5)}
   {\psset{linewidth=0.05}\psline(17.5,5)(17.5,6)}   
   {\psset{linewidth=0.05}\psline(15.5,5)(15.5,6)}
   {\psset{linewidth=0.05}\psline(16.5,5)(16.5,6)}
   
   {\psset{linewidth=0.05}\psline(17.5,6)(18,5)}
   {\psset{linewidth=0.05}\psline(18,6)(17.5,5)}
   {\psset{linewidth=0.05}\psline(18,6)(18,5)}

   \rput(16,4.5){\small(mixed)}
\end{pspicture}
\caption{To compute the prefix $\psi_1^3$ of $|\psi\rangle:=\frac 1 {\sqrt{2}}\left(|00\rangle-|1111\rangle\right)$, the qubit string
is embedded in the tape Hilbert space $\hr_\N$. The result is mixed.}
\label{FigTM}
\end{center}
\end{figure}

The tape of a quantum Turing machine (QTM) \cite{BernsteinVazirani} is very similar, but it can
also carry superpositions and mixtures of classical tape configurations. Our idea is to embed qubit
strings in the tape's Hilbert space, use the operations (partial trace or
tensor product) on this Hilbert space, and then ``read off'' the result to get back to string space.
This is illustrated in Figure~\ref{FigTM}.

Since we are not interested in the
details of the computation, but only want to embed qubit strings into the tape, it is sufficient for our purpose
to have one-way infinite tapes, i.e. tapes with cells indexed by the positive natural numbers.
We can then define $\hr_\N$ as the Hilbert space which
is spanned by the classical tape configurations $T_\N$, where
\[
   T_\N:=\{t\in\{0,1,\#\}^\N\,\,|\,\, \mbox{ only finitely many }t_i\neq\#\}.
\]
For example, $|101\#0\#\#\#\ldots\rangle\in\hr_\N$, but $|111\ldots\rangle\not\in\hr_\N$.
Since there are only countably many configurations with finitely many non-blanks, the
Hilbert space $\hr_\N$ is separable.

The set of classical configurations $T$
has a product structure: For any subset $I\subset\N$, define
\[
   T_I:=\{t\in\{0,1,\#\}^I\,\,|\,\, \mbox{only finitely many }t_i\neq 0\}.
\]
Then, $T_I$ gives us all the possibilities to choose symbols from $\{0,1,\#\}$ at the cells indexed in $I$.
If $I\subset\N$ is a finite set, then $T_I=\{0,1,\#\}^I$. Moreover, it holds
$T_\N=T_I\times T_{\N\setminus I}$.
Thus, if $\hr_I$ is the Hilbert space spanned by the classical configurations in $T_I$, then $\hr_\N$
has the tensor product structure
\[
   \hr_\N=\hr_I\otimes\hr_{\N\setminus I}.
\]

Writing down a configuration in $T_I$ symbol by symbol, we get strings like $s=01\#0\#1\#\#\ldots$
or $t=1001\#\#\#\ldots$. We call configurations like $t$ (but not like $s$) that start with some bits, followed by blanks,
{\em bit string configurations}. The bit string configurations in $T_I$ span a Hilbert subspace $\hr_I^S\subset\hr_I$.
For example, if $I=\{1,2\}$, then $\hr_I$ is a $9$-dimensional subspace of $\hr_\N$, containing the strings
$00$, $01$, $0\#$, $10$, $11$, $1\#$, $\#0$, $\#1$, $\#\#$ as orthonormal basis vectors.
In contrast, $\hr_I^S$ is then $7$-dimensional, and does not contain the strings $\#0$ and $\#1$, because
they are not bit string configurations.

If $s$ is a classical bit string with $\ell(s)\leq |I|$, then we can embed the vector $|s\rangle\in\hr_\s$
into $\hr_I^S$: just define
\[
   U_I|s\rangle:=|\underbrace{s}_{[i_1,\ldots,i_{\ell(s)}]}\underbrace{\#\#\#\ldots}_{[i_{\ell(s)+1},i_{|I|}]}\rangle,
\]
i.e. fill up the cells with the bits of $s$, followed by blanks,
and linearly extend this map to the other vectors. Then $U_I$ is a unitary map of the form
\[
   U_I:\hr_{\leq |I|}\to \hr_I^S,
\]
i.e. it embeds all quantum bit strings of base length less than $|I|$ unitarily into $\hr_I^S$.

To embed a qubit string into the tape Hilbert space, we first use the maps $U_I$ to map
it into $\hr_I^S$, and then we would like to treat the resulting state as a state on $\hr_I$.
The map which embeds vectors from $\hr_I^S$ into $\hr_I$ will be called $\iota_I$.
This is a linear map $\iota_I:\hr_I^S\to\hr_I$ which does not affect vectors at all,
i.e. $\iota_I |\varphi\rangle=|\varphi\rangle$ for all $|\varphi\rangle\in\hr_I^S$.
However, note that its adjoint $\iota_I^*$ is not so trivial: in fact, it projects vectors
from $\hr_I$ onto $\hr_I^S$, that is,
\[
   \iota_I^*\iota_I=\idn_{\hr_I^S},\mbox{ but }\iota_I \iota_I^*=\mathbb{P}(\hr_I^S).
\]
This reflects the fact that only elements from $\hr_I^S$ can be treated as valid qubit strings;
elements from the orthogonal complement have to be dismissed somehow (we will see in some examples
later how this projection property enters calculations).

\begin{figure}[!hbt]
\[
\begin{xy}
  \xymatrixrowsep{0pc}\xymatrix{
      \hr_\s \ar@<2pt>@^{->}[r]^-{U_I}  & \hr_I^S\ar@<2pt>@^{->}[l]^-{U_I^*} \ar@/^/[rr]^-{\iota_I}  
      & \subset  & \hr_I
      \ar@/^/[ll]^-{\iota_I^*} \\
      \| &  \bigcap & & \bigcap\\
      \hr_\s \ar@<2pt>@^{->}[r]^-{U_\N}  & \hr_\N^S\ar@<2pt>@^{->}[l]^-{U_\N^*} \ar@/^/[rr]^-{\iota_\N} &\subset & \hr_\N
      \ar@/^/[ll]^-{\iota_\N^*}  \ar@/_1pc/[uu]_-{\Tr_{\N\setminus I}} \ar@/^1pc/[dd]^-{\Tr_{\N\setminus J}}\\
      \bigcup & \bigcup & & \bigcup \\
      \hr_{\leq |J|} \ar@<2pt>@^{->}[r]^-{U_J}  & \hr_J^S\ar@<2pt>@^{->}[l]^-{U_J^*} \ar@/^/[rr]^-{\iota_J}  
      & \subset  & \hr_J
      \ar@/^/[ll]^-{\iota_J^*} \\
  }
\end{xy}
\]
\caption{The structure of maps  and spaces used to embed qubit strings from $\hr_\s$ into the tape Hilbert space $\hr_\N$.
Here, $I\subset \N$ is assumed to be infinite, and $J\subset\N$ is finite.}
\label{FigStructure}
\end{figure}
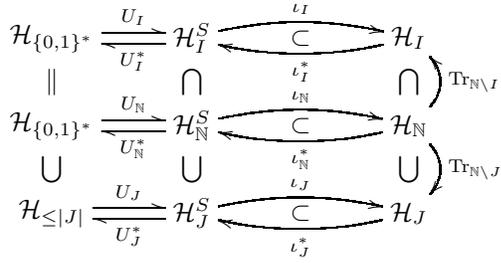

Figure~\ref{FigStructure} shows the structure of subspaces and maps that we have just defined.

We can now use this tensor product structure in $\hr_\N$ to define tensor products and partial traces (prefixes)
on $\hr_\s$.

\begin{definition}[Prefixes and Restrictions of Qubit Strings]
For every qubit string $\rho\in\St(\hr_\s)$ and $I\subset\N$, we define the
restriction $\rho_I$ of $\rho$ to the bit positions $I$ as
\[
   \rho_I:=U_I^* \iota_I^*\Tr_{\N\setminus I}\left(\iota_{\N} U_\N \rho U_\N^*\iota_{\N}^*\right)\iota_I U_I.
\]
We also use the notation $\rho_m^n:=\rho_{[m,n]}$ for $m,n\in\N$, and we call $\rho^n:=\rho_1^n$ the
{\em $n$-qubit prefix} of $\rho$.
\end{definition}

While this definition looks quite formal, it is easy to implement in calculations: just fill in ``missing'' qubits
with blanks ($\#$) that are treated as orthogonal to $0$ or $1$. For example, if $|\psi\rangle:=\frac 1 {\sqrt{2}}
\left(|1\rangle+|110\rangle\right)$, then
\begin{eqnarray*}
   \psi_1^2&:=&|\psi\rangle\langle\psi|_1^2=\frac 1 2\left(\strut|1\#\#\rangle\langle 1\#\#|+
   |1\#\#\rangle\langle 110|
   \right.\\
   && \left.\strut\qquad\qquad\quad \,\,+|110\rangle\langle 1\#\#|+|110\rangle\langle 110|\right)_1^2\\
   &=&\frac 1 2 |1\#\rangle\langle 1\#|+\frac 1 2 |1\#\rangle\langle 11|\cdot\Tr |\#\rangle\langle 0|\\
   && + \frac 1 2 |11\rangle\langle 1\#|\cdot \Tr |0\rangle\langle\#| + \frac 1 2 |11\rangle\langle 11|\\
   &=&\frac 1 2 |1\rangle\langle 1| + \frac 1 2 |11\rangle\langle 11|.
\end{eqnarray*}
This also shows that prefixes of pure qubit strings do not have to be pure, in contrast to the classical situation.

For every density operator $\rho$, the restriction or fragment $\rho_I$ is also a valid density operator.
This follows from the fact that the partial trace $\Tr_{\N\setminus I}\left(\iota_\N U_\N \rho U_\N^* \iota_\N^*\right)$
maps $\rho$ into the states on the subspace $\hr_I^S$, i.e. the resulting density operator
has vanishing support
on non-bit string configurations like $1\#0$.
In general, if $\sigma$ is a density operator with full support on $\hr_\N^S$, then $\Tr_{\N\setminus I}\sigma$
is a density operator with full support on $\hr_I^S$.
This is easy to check; we omit the simple proof.

We can also define a tensor product on $\hr_\s$. The idea is to define some product $A\otimes_I B$
which means that we put the observable $A$ in the places indexed by $I$, and the observable $B$ in all
the other places. However, this tensor product does not make quantum mechanically sense for all choices of factors and subsets;
we will discuss this problem afterwards.
\begin{definition}[Tensor Product on String Space]
\lineclear
Let $I\subset\N$ with $n:=|I|$ and $m:=|\N\setminus I|$ (possibly n=$\infty$ or $m=\infty$). If $A\in\mathcal{B}(\hr_{\leq n})$
and $B\in\mathcal{B}(\hr_{\leq m})$ are operators on the string space, then we define the tensor product
$A\otimes_I B$ as
\begin{eqnarray*}
   U_\N^* \iota_\N^*\left[ \left(\iota_I U_I A U_I^* \iota_I^*\right)\otimes 
   \left(\iota_{\N\setminus I} U_{\N\setminus I} B U_{\N\setminus I}^* \iota_{\N\setminus I}^*\right)\right]
   \iota_\N U_\N.
\end{eqnarray*}
If $\rho$ and $\sigma$ are qubit strings such that $\rho$ is a length eigenstate, we
set
\[
   \rho\otimes\sigma:=\rho\otimes_{[1,\ell(\rho)]}\sigma.
\]
\end{definition}

If $\rho$ is a length eigenstate, the tensor product is well-behaved. For example,
\[
   \left(\rho\otimes\sigma\right)^{\ell(\rho)}=\rho\mbox{ and }
   \left(\rho\otimes\sigma\right)_{\ell(\rho)+1}^{\ell(\rho)+\ell(\sigma)}=\sigma.
\]
This can easily be checked by inserting the definition.
In this expression, we can replace $\ell(\rho)+\ell(\sigma)$ by any larger integer or
by $\infty$, but not by $\ell(\rho)+\bar\ell(\sigma)$. For classical strings,
the tensor product turns into concatenation: Let $n:=\ell(s)$, then
\begin{eqnarray}
   |s\rangle\otimes|t\rangle&=&U_\N^* \iota_\N^*\left(\iota_{[1,n]}U_{[1,n]}|s\rangle\otimes 
   \iota_{[n+1,\infty)} U_{[n+1,\infty)} |t\rangle\right)\nonumber\\
   &=& U_\N^* \iota_\N^*\left(|s\rangle\otimes |t\#\#\ldots\rangle\right)=U_\N^* \iota_\N^* |s\circ t\#\#\ldots\rangle\nonumber\\
   &=&|s\circ t\rangle.\label{EqConcatClass}
\end{eqnarray}
However, we need the restriction that $\rho$ shall be a length eigenstate to assure that the intermediately computed
density operator lives on the subspace $\hr_\N^S$, i.e. can be interpreted as a valid qubit string.
Otherwise, we get into trouble and lose normalization. For example, if $|\psi\rangle=\frac 3 5 |\lambda\rangle+\frac 4 5 |0\rangle$ (recall
that $\lambda$ is the empty string), then the unphysical tensor product $|\psi\rangle\otimes|1\rangle:=|\psi\rangle
\otimes_{\{1\}} |1\rangle$ is
\begin{eqnarray*}
   |\psi\rangle\otimes|1\rangle&=&U_\N^* \iota_\N^*\left(\iota_{\{1\}}U_{\{1\}}|\psi\rangle\otimes 
   \iota_{[2,\infty)} U_{[2,\infty)} |1\rangle\right)\\
   &=&U_\N^* \iota_\N^*\left(\left(\frac 3 5 |\#\rangle+\frac 4 5|0\rangle\right)\otimes |1\#\#\ldots\rangle\right)\\
   &=&\frac 3 5 U_\N^* \underbrace{\iota_\N^*|\# 1 \#\#\ldots\rangle}_0+\frac 4 5 U_\N^* \iota_\N^*|01\#\#\ldots\rangle\\
   &=&\frac 4 5 |01\rangle,
\end{eqnarray*}
which is no state vector, because it is not normalized.

This reflects the fact that is is physically
unfeasible to put the states $|\psi\rangle$ and $|1\rangle$ side by side on the tape without intermediate blanks.
A similar situation arises when we use the tensor product $\otimes_I$ for some bad choice of $I$:
For example, $|11\rangle\otimes_{[3,4]} |0\rangle = U_\N^* \iota_\N^* |0\# 1 1 \#\#\ldots\rangle=0$.

Apart from the fact that bad choices of $I$ and/or $\rho$ destroy normalization,
the tensor product $\otimes_I$ behaves
very much like the ``classical'' tensor product on a bipartite Hilbert space. For example, we have the following
relation between the partial trace and the tensor product:
\begin{lemma}\label{LemRestrictionTensProd}
If $\rho$ is a qubit string and $I\subset \N$, then
\[
   \Tr (\rho_I A)=\Tr(\rho A\otimes_I \idn)
\]
for every $A\in\mathcal{B}(\hr_{\leq |I|})$.
\end{lemma}
\proof
This is a matter of calculation:
\begin{eqnarray*}
\Tr(\rho_I A)&=&\Tr\left[ U_I^* \iota_I^* \Tr_{\N\setminus I}\left(\iota_\N U_\N \rho U_\N^* \iota_\N^*\right)\iota_I U_I A\right]\\
&=&\Tr\left[ \Tr_{\N\setminus I} \left(\iota_\N U_\N \rho U_\N^* \iota_\N^*\right) \iota_I U_I A U_I^* \iota_I^*\right]\\
&=&\Tr\left[ \left(\iota_\N U_\N \rho U_\N^* \iota_\N^*\right)\cdot \left( ( \iota_I U_I A U_I^* \iota_I^*) \otimes \idn_{\hr_{\N\setminus I}}
\right)\right]\\
&=&\Tr\left[ \rho U_\N^* \iota_\N^* \left( (\iota_I U_I A U_I^* \iota_I^*)\otimes \idn_{\hr_{\N\setminus I}}\right) \iota_\N U_\N\right].
\end{eqnarray*}
At this point, we can split the operator $\idn_{\hr_{\N\setminus I}}$ into two parts: It holds
\[
   \idn_{\hr_{\N\setminus I}}=\mathbb{P}\left(\hr_{\N\setminus I}^S \right)+\mathbb{P}\left(\left(\hr_{\N\setminus I}^S\right)^\perp\right)
   =:\mathbb{P}+\mathbb{P}^\perp.
\]
We would like to show that the part of the above expression which corresponds to $\mathbb{P}^\perp$ vanishes.
Let $\tilde A:=\iota_I U_I A U_I^* \iota_I^*$. Let $|t\rangle\in\hr_\N$ be arbitrary and let $|s\rangle\in\hr_\N^S$ be a
bit string configuration. Denote $\N\setminus I=:\{j_1,j_2,j_3,\ldots\}$, then
\begin{eqnarray*}
   \langle s|\iota_\N^*\left(\tilde A\otimes \mathbb{P}^\perp\right)|t\rangle&=&
   \langle s|\left(\tilde A\otimes \mathbb{P}^\perp\right)|t\rangle\\
   &=&\langle \cdot|\tilde A|\cdot\rangle\cdot
   \langle s_{j_1} s_{j_2}\ldots|\mathbb{P}^\perp | t_{j_1} t_{j_2}\ldots\rangle.
\end{eqnarray*}
If $t_{j_1}t_{j_2}\ldots$ is a bit string configuration then $\mathbb{P}^\perp |t_{j_1} t_{j_2}\ldots\rangle=0$.
Otherwise, $\mathbb{P}^\perp |t_{j_1} t_{j_2}\ldots\rangle=|t_{j_1} t_{j_2}\ldots\rangle$; but then, the scalar
product is zero, since $s_{j_1} s_{j_2}\ldots$ is a bit string configuration.
Consequently,
\[
   \iota_\N^*\left(\tilde A \otimes \mathbb{P}^\perp\right)=0.
\]
Thus, we get
\begin{equation}
   \Tr(\rho_I A)=\Tr\left[\rho U_\N^* \iota_\N^*\left( (\iota_I U_I A U_I^* \iota_I^* )\otimes \mathbb{P}\right) \iota_\N U_\N\right].
   \label{EqRes}
\end{equation}
Finally, inserting
\[
   \mathbb{P}=\iota_{\N \setminus I} \iota_{\N \setminus I}^* = \iota_{\N\setminus I} U_{\N\setminus I}
   \idn_{\hr_{\leq |\N\setminus I|}} U_{\N\setminus I}^* \iota_{\N\setminus I}^*,
\]
we see that the right-hand side of (\ref{EqRes}) equals $\Tr(\rho A\otimes_I \idn)$.
\qed

Our aim is to define prefix-free quantum codes. For this reason, we have to define what we mean by concatenation
of quantum bit strings. This can be done by linear extension of the classical concatenation operation:
\begin{definition}[Concatenation of Quantum Bit Strings]
\label{DefQConcat}
For every qubit string $|\psi\rangle=\sum_{t\in\s}\alpha_t |t\rangle$ and $s\in\s$, we define the concatenation
$|\psi\circ s\rangle$ by
\[
   |\psi\circ s\rangle:=\sum_{t\in\s} \alpha_t |t\circ s\rangle.
\]
Moreover, if $|\varphi\rangle=\sum_{t\in\s}\beta_t |t\rangle$ is another qubit string,
then we set
\[
   |\psi\circ \varphi\rangle:=|\psi\rangle\circ |\varphi\rangle:=\sum_{t\in\s}\beta_t |\psi\circ t\rangle.
\]
\end{definition}
Clearly, for every fixed $s\in\s$, the map $|\psi\rangle\mapsto |\psi\circ s\rangle$ is an isometry.
However, the map $|\psi\rangle\mapsto |\psi\circ \varphi\rangle$ is not isometric in general. For example,
if $|\psi\rangle=\frac 1 {\sqrt{2}} \left(|0\rangle+|00\rangle\right)$ and
$|\varphi\rangle=\frac 1 {\sqrt{2}} \left(|0\rangle-|00\rangle\right)$, then
\begin{eqnarray*}
   |\psi\circ\varphi\rangle&=&\left(\frac 1 {\sqrt{2}}|0\rangle+\frac 1 {\sqrt{2}} |00\rangle\right)\circ
   \left(\frac 1 {\sqrt{2}}|0\rangle-\frac 1 {\sqrt{2}} |00\rangle\right)\\
   &=&\frac 1 2 |0\circ 0\rangle-\frac 1 2 |0\circ 00\rangle + \frac 1 2 |00\circ 0\rangle
   -\frac 1 2 |00\circ 00\rangle\\
   &=&\frac 1 2 |00\rangle - \frac 1 2 |0000\rangle.
\end{eqnarray*}
This resulting vector is not normalized. Thus, the concatenation operation can be unphysical if the first
quantum bit string is not a length eigenstate. But if it is, it coincides with the tensor product. We omit the simple proof.
\begin{lemma}[Concatenation and Tensor Product]
\label{LemConcatTensProd}
\lineclear
If $|\psi\rangle,|\varphi\rangle\in\hr_\s$ are qubit strings such that $|\psi\rangle$ is a length eigenstate,
then $|\psi\rangle\circ|\varphi\rangle=|\psi\rangle\otimes|\varphi\rangle$.
\end{lemma}

\section{Prefix-Free Quantum Bit Strings and the Quantum Kraft Inequality}
\label{SecIntuitive}
Now we are ready to define prefix-free sets of quantum bit strings.
\begin{definition}[Prefix-Free Sets of Qubit Strings]
\label{DefPrefixFree}
\lineclear
A set $M\subset \hr_\s$ of qubit strings is called {\em prefix-free}, if one of the four
following equivalent conditions holds:
\begin{itemize}
\item[(1)] For every $|\varphi\rangle,|\psi\rangle\in M$ and classical string $s\in\s\setminus\{\lambda\}$, it holds
$\langle \varphi|\psi\circ s\rangle=0$.
\item[(2)] For every $|\varphi\rangle,|\psi\rangle\in M$ and qubit string $|\chi\rangle\perp |\lambda\rangle$, it holds
$\langle \varphi|\psi\circ\chi\rangle=0$.
\item[(3)] For every $|\varphi\rangle,|\psi\rangle\in M$ and classical strings $s,t\in\s$ with $s\neq t$, it holds
$\langle \varphi \circ t|\psi\circ s\rangle=0$.
\item[(4)] For every $|\varphi\rangle,|\psi\rangle\in M$ and qubit strings $|\chi\rangle,|\tau\rangle\in\hr_\s$ with $|\chi\rangle
\perp|\tau\rangle$, it holds
$\langle \varphi\circ\tau|\psi\circ\chi\rangle=0$.
\end{itemize}
\end{definition}
It is easy to see that these four conditions are equivalent: $(2)\Rightarrow(1)$, $(3)\Rightarrow(1)$ and $(4)\Rightarrow(3)$ are
trivial. $(1)\Rightarrow(2)$
and $(3)\Rightarrow(4)$ follow by expanding the definition of $|\psi\circ\chi\rangle$ and $|\varphi\circ\tau\rangle$.
To see that $(1)\Rightarrow(3)$, note that for every $|\varphi\rangle,|\psi\rangle\in\hr_\s$, it holds
\[
   \langle \varphi\circ 0 | \psi\circ 1\rangle=\langle\varphi\circ 1 |\psi\circ 0 \rangle=0
\]
and also
\[
   \langle \varphi\circ 0 | \psi\circ 0\rangle=\langle\varphi\circ 1 |\psi\circ 1 \rangle=\langle \varphi|\psi\rangle.
\]
It follows for $s,t\in\s$ with $\ell(s)\geq \ell(t)$ that
\[
   \langle\varphi\circ t | \psi\circ s\rangle=\left\{
      \begin{array}{cl}
         \langle \varphi|\psi\circ s_1^{\ell(s)-\ell(t)} \rangle & \mbox{if }t=s_{\ell(s)-\ell(t)+1}^{\ell(s)}\\
         0 & \mbox{otherwise.}
      \end{array}
   \right.
\]
The most interesting case is when the set $M$ is itself a closed subspace $\hr$ of $\hr_\s$. Such a ``prefix-free Hilbert
space'' is the quantum analogue of a (prefix-free) code book. For this situation, we have the following lemma:

\begin{lemma}
\label{LemPrefixFreeHR}
A Hilbert space $\hr\subset \hr_{\{0,1\}^*}$ is prefix-free if and only if it has a prefix-free orthonormal basis.
In this case, {\em every} orthonormal basis of $\hr$ is prefix-free.
\end{lemma}
\proof
Let $\{e_i\}_i$ be a prefix-free orthonormal basis of $\hr$ and let $|v\rangle,|w\rangle\in\hr$ and $s\neq \lambda$
be arbitrary. Expanding $|w\rangle$ as $|w\rangle=\sum_j \alpha_j |e_j\rangle$, we get
\[
   \langle e_i\circ s|w\rangle=\sum_j \alpha_j \underbrace{\langle e_i\circ s|e_j\rangle}_0=0.
\]
Moreover, if $|v\rangle=\sum_i \beta_i |e_i\rangle$, then
\[
   \langle v\circ s|w\rangle=\sum_i \beta_i^* \underbrace{\langle e_i\circ s|w\rangle}_0 = 0.
   \qquad\qquad\qquad\qquad\qquad\,\,\,\mbox{\qed}
\]
Unlike in the classical case, quantum bit strings can be prefixed by themselves. In other words, there
are qubit strings $|\varphi\rangle\in\hr_\s$ such that the set $\{|\varphi\rangle\}$ is not prefix-free.
For example, if $|\varphi\rangle:=\frac 1 {\sqrt{2}}\left(|\lambda\rangle+|0\rangle\right)$, then
$\langle\varphi|\varphi\circ 0\rangle=\frac 1 2 \neq 0$.

However, even if a qubit string consists of a superposition of classical strings where some string is the
prefix of another, it need not be a prefix of itself. For example, the qubit string
\[
   |\varphi\rangle:=\frac 1 2 |1\rangle+\frac 1 2 |10\rangle + \frac 1 2 |0\rangle-\frac 1 2 |00\rangle
\]
is not a prefix of itself. In particular, $\langle\varphi|\varphi\circ 0\rangle=0$.

So far, the definition of a prefix-free Hilbert space has been a purely formal generalization of the classical
definition. Now we will see that this definition really has desirable properties: every basis vector of length $n$
can be distinguished with certainty from every other (even longer) basis vector by measurement on the first
$n$ qubits only. Yet, this is only true for orthonormal bases of length eigenstates.
\begin{lemma}
\label{LemDistinguish}
An orthonormal system $M\subset\hr_\s$ which consists entirely of length eigenstates is prefix-free if and only if
for every $|\varphi\rangle,|\psi\rangle\in M$ with $|\varphi\rangle\neq |\psi\rangle$, it holds
\[
   \langle \psi | \varphi^{\ell(\psi)} |\psi\rangle=0.
\]
\end{lemma}
\proof
Let $|\varphi\rangle,|\psi\rangle\in M$ with $|\varphi\rangle\neq |\psi\rangle$ and let $n:=\ell(\psi)$. Using
Lemma~\ref{LemRestrictionTensProd} and Lemma~\ref{LemConcatTensProd}, we get
\begin{eqnarray*}
\langle \psi|\left(|\varphi\rangle\langle\varphi|_1^n\right)|\psi\rangle
&=&\Tr\left(|\psi\rangle\langle\psi|\cdot |\varphi\rangle\langle\varphi|_1^n\right)\\
&=&\Tr\left(|\varphi\rangle\langle\varphi|\cdot |\psi\rangle\langle\psi|\otimes\idn   \right)\\
&=&\sum_{s\in\s}\langle \varphi| \left(\strut|\psi\rangle\langle\psi|\otimes|s\rangle\langle s|\right)|\varphi\rangle\\
&=&\sum_{s\in\s} \langle\varphi|\psi\circ s\rangle\langle\psi\circ s|\varphi\rangle\\
&=&\sum_{s\in\s\setminus\{\lambda\}} |\langle\varphi|\psi\circ s\rangle|^2.
\end{eqnarray*}
Thus, the left-hand side is zero if and only if every addend on the right-hand side is zero. For the case that $|\varphi\rangle=|\psi\rangle$,
note that a length eigenstate can never be a prefix of itself.
\qed
This lemma also shows that in case there exists an orthonormal basis of length eigenstates, our concept
of a prefix-free Hilbert space is equivalent to the definition by Schumacher and Westmoreland~\cite{SchumacherWestmoreland}.
However, we do not fill up the code words with zeroes, and we have no global upper bound on the code word length.

Moreover, we can study ``quantum prefix code books'', that is prefix-free Hilbert spaces, that do not have any
basis of length eigenstates. The orthonormal bases of such Hilbert spaces do not necessarily have the
property of Lemma~\ref{LemDistinguish}. Here is an example:
\begin{example}[Strange Prefix-Free Hilbert Space]
\label{ExStrange}
Let $|\psi\rangle,|\varphi\rangle\in\hr_\s$ be the two vectors
\[
   M:=\left\{
      \frac 1 {\sqrt{2}}\left(|1\rangle+|01\rangle\right), \frac 1 {\sqrt{2}} \left(|10\rangle-|010\rangle\right)
   \right\}.
\]
Then $M$ is prefix-free. In particular, it holds
\[
   \langle\varphi|\psi\circ 0\rangle=\frac 1 2 \left(\langle 10|-\langle 010|\right)\left(|10\rangle+|010\rangle\right)=0.
\]
Moreover, $M$ is an orthonormal system. According to Lemma~\ref{LemPrefixFreeHR}, $M$ spans a prefix-free Hilbert space.
This Hilbert space does {\em not} possess an orthonormal basis of length eigenstates. Also, $|\varphi\rangle$ and $|\psi\rangle$
can {\em not} be distinguished by looking at the first $2$ qubits only: It holds
\[
   \varphi^{\ell(\psi)}=|\varphi\rangle\langle\varphi|_1^2 = \frac 1 2 |10\rangle\langle 10| + \frac 1 2 |01\rangle\langle 01|,
\]
so $\langle\psi|\varphi^{\ell(\psi)}|\psi\rangle=\frac 1 4 \neq 0$.
\end{example}

We would like to prove a quantum version of the Kraft inequality. This has been done by Schumacher and
Westmoreland~\cite{SchumacherWestmoreland}
for the case that the code space is spanned by length eigenstates. However, we would like to prove a generalization of that inequality
for arbitrary prefix-free Hilbert spaces, even if they do not have an orthonormal basis of length eigenstates. This needs some preparation:

\begin{proposition}
\label{PropKraft}
Let $\{|e_i\rangle\}_{i\in I}\subset\hr_\s$ be any orthonormal system, spanning a Hilbert space $\hr\subset\hr_\s$.
Then, we have
\[
   \sum_{i\in I} 2^{-\ell(e_i)} \leq \sum_{i\in I} 2^{-\bar\ell(e_i)} \leq \Tr\left(2^{-\Lambda} \mathbb{P}(\hr)\right).
\]
If the left-most expression is finite, then equality holds if and only if every $|e_i\rangle$ is a length eigenstate.
\end{proposition}
\proof
The first inequality is trivial, since $\bar\ell(\varphi)\leq\ell(\varphi)$ for every $|\varphi\rangle\in\hr_\s$.
Every vector can be expanded as
\[
   |\varphi\rangle=\sum_{l=0}^\infty \sum_{k=1}^{2^l} \alpha_{kl} |s_{kl}\rangle\qquad\mbox{with }s_{kl}\in\{0,1\}^l.
\]
It follows that
\[
   \bar\ell(\varphi)=\langle\varphi|\Lambda|\varphi\rangle=\sum_{l=0}^\infty \sum_{k=1}^{2^l} l |\alpha_{kl}|^2.
\]
Since the function $x\mapsto 2^{-x}$ is convex, it holds for every set of probabilities $\{\lambda_i\}_i$ with
$\sum_i \lambda_i=1$ and $\lambda_i\geq 0$ that $2^{-\sum_i \lambda_i a_i} \leq \sum_i \lambda_i 2^{-a_i}$, so
\[
   2^{-\bar\ell(\varphi)}=2^{-\sum_l \sum_k |\alpha_{kl}|^2 l}
   \leq \sum_{l=0}^\infty \sum_{k=1}^{2^l}|\alpha_{kl}|^2 2^{-l}=\langle\varphi|2^{-\Lambda}|\varphi\rangle.
\]
The second inequality in the proposition follows then from inserting $e_i$ for $\varphi$ and summing over $i$. This also
shows that left- and right-hand side can only be equal if all the addends are equal, that is, if
\begin{equation}
   2^{-\ell(e_i)}=\langle e_i|2^{-\Lambda}|e_i\rangle
   \label{EqGleich}
\end{equation}
for every $i$. But $\langle e_i|2^{-\Lambda}|e_i\rangle=\sum_{l,k} |\alpha_{kl}|^2 2^{-l}$ is a convex combination
of different $2^{-l}$-terms, where the largest term is $2^{-\ell(e_i)}$. Thus, if (\ref{EqGleich}) holds, then all the weight must be concentrated
where $l=\ell(e_i)$, such that $e_i$ is a length eigenstate.\qed

Now we are ready to state the quantum generalization of the Kraft inequality:
\begin{theorem}[Quantum Kraft Inequality]
\label{TheKraft}
\lineclear
Let $\{|e_i\rangle\}_{i\in I}\subset\hr_\s$ be a prefix-free orthonormal system, spanning a Hilbert space $\hr\subset\hr_\s$.
Then, it holds
\[
   \sum_{i\in I} 2^{-\ell(e_i)} \leq \sum_{i\in I} 2^{-\bar\ell(e_i)} \leq \Tr\left(2^{-\Lambda} \mathbb{P}(\hr)\right)\leq 1.
\]
Equality holds for the left three terms if and only if every $|e_i\rangle$ is a length eigenstate.
\end{theorem}
\proof
Using Proposition~\ref{PropKraft}, all we have to prove is that $\Tr\left(2^{-\Lambda} \mathbb{P}(\hr)\right)\leq 1$.
We may restrict to the case that $I$ is a finite set, and that the base lengths of all $|e_i\rangle$ are finite. The general
case then follows from a continuity argument.

It will be helpful for the proof to define a quantity called the {\em weight} of a vector $|\varphi\rangle\in\hr_\s$
on some classical string $s\in\s$. We define it as
\[
   w_\varphi(s):=\sum_{t\in\s} |\langle \varphi\circ t|s\rangle|^2.
\]
We first show that $\sum_{i\in I} w_{e_i}(s)\leq 1$, due to the prefix property of $\{|e_i\rangle\}$. If $i\neq j$ or $u\neq t$,
then $\langle e_i \circ t|e_j \circ u\rangle=0$. Thus, the set
\[
   \left\{|e_i\circ t\rangle\,\,|\,\, i\in I, t\in\s\right\}
\]
is an orthonormal system in $\hr_\s$. It follows that
\[
   \sum_{i\in I} \sum_{t\in\s} |e_i\circ t\rangle\langle e_i \circ t|\leq\idn.
\]
In particular, it holds
\begin{eqnarray*}
1=\langle s|s\rangle &\geq& \sum_{i\in I} \sum_{t\in\s} \langle s|e_i\circ t\rangle\langle e_i\circ t|s\rangle\\
&=&\sum_{i\in I} \sum_{t\in\s} |\langle e_i\circ t|s\rangle|^2
=\sum_{i\in I}w_{e_i}(s).
\end{eqnarray*}
Now choose some $n>\max_{i\in I} \ell(e_i)$. Clearly, the ``full weight'' $W$ on $\{0,1\}^n$ satisfies
\[
   W:=\sum_{u\in\{0,1\}^n} \sum_{i\in I} w_{e_i}(u)\leq |\{0,1\}^n|\cdot 1 = 2^n.
\]
Expand each $|e_i\rangle$ as $|e_i\rangle=\sum_{s\in\s} \alpha_{is} |s\rangle$. In fact, if $\ell(s)>n$, then
$\alpha_{is}=0$ for every $i$. If $s,t\in\s$, then
\[
   \langle e_i\circ t | u\rangle=\sum_{s\in\s} \alpha_{is}^* \underbrace{\langle s\circ t |u\rangle}_{\delta_{s\circ t,u}},
\]
so $|\langle e_i\circ t|u\rangle|^2=|\alpha_{is}|^2$ for that $s$ for which $s\circ t=u$ (or zero if there is no
such $s$). Thus,
\[
   w_{e_i}(u)=\sum_{t\in\s}|\langle e_i\circ t|u\rangle|^2=\sum_{s\mbox{ prefix of }u} |\alpha_{is}|^2.
\]
This statement can also be understood as follows: For every $s$, the vector $|e_i\rangle$ contributes to the weight of
every extension $u\in\{0,1\}^n$ of $s$ with $|\alpha_{is}|^2$. There are $2^{n-\ell(s)}$ such extensions. Thus, $|e_i\rangle$
contributes to the full weight $W$ some weight $2^{n-\ell(s)}|\alpha_{is}|^2$ for every $s$.
Summing over $s$ and then over $i$ yields
\begin{eqnarray*}
2^n &\geq& W \geq \sum_{i\in I} \sum_{s\in\s} 2^{n-\ell(s)} |\alpha_{is}|^2\\
&=&2^n \sum_{i\in I} \langle e_i|2^{-\Lambda}|e_i\rangle=2^n \Tr\left(2^{-\Lambda} \mathbb{P}(\hr)\right).
\qquad\qquad\,\,\mbox{\qed}
\end{eqnarray*}

\begin{example}Suppose the Hilbert space $\hr$ is spanned by the orthonormal basis vectors
\[
   \left\{
      \underbrace{\frac 1 {\sqrt{2}}\left(|1\rangle+|01\rangle\right)}_{|e_1\rangle},
      \underbrace{\frac 1 {\sqrt{2}} \left(|10\rangle-|010\rangle\right)}_{|e_2\rangle},
      \underbrace{|00\rangle}_{|e_3\rangle}
   \right\},
\]
which is the basis from Example~\ref{ExStrange}, plus the basis vector $|00\rangle$. Since this
orthonormal basis is prefix-free, so is $\hr$ according to Lemma~\ref{LemPrefixFreeHR}. The quantum Kraft
inequality from Theorem~\ref{TheKraft} can be confirmed by explicit calculation:
\begin{eqnarray*}
   \sum_{i=1}^3 2^{-\ell(e_i)}&=&2^{-2}+2^{-3}+2^{-2}=\frac 5 8 =0.625,  \\
   \sum_{i=1}^3 2^{-\bar\ell(e_i)}&=&2^{-\frac 3 2}+2^{-\frac 5 2}+2^{-2}=\frac{2+3\sqrt{2}}{8}\approx 0.78,\\
   \Tr\left(2^{-\Lambda}\mathbb{P}(\hr)\right)&=&\sum_{i=1}^3 \langle e_i|2^{-\Lambda}|e_i\rangle=\frac{13}{16}=0.8125.
\end{eqnarray*}
\end{example}

\section{Summary and Perspectives}
We have defined prefixes, restrictions, tensor products, and concatenation of indeterminate-length qubit strings.
This led to a very general definition of prefix-free Hilbert spaces $\hr$ (Definition~\ref{DefPrefixFree}).
In the special case that $\hr$ has a basis of length eigenstates, this definition is equivalent to that by
Schumacher and Westmoreland (Lemma~\ref{LemDistinguish}). Consequently, these Hilbert spaces can be used for
lossless quantum compression as shown in \cite{SchumacherWestmoreland}.

However, there are strange prefix-free Hilbert spaces (cf. Example~\ref{ExStrange}) that do {\em not} have the nice
property that different code words can be distinguished with certainty by looking at their prefixes only.
Still, these spaces obey a generalized quantum Kraft inequality (Theorem~\ref{TheKraft}).

It is an interesting open
problem if
these spaces are useful for quantum compression.
The key question is whether the concatenation of qubit strings makes physically sense; if it does, it can
serve as a replacement for the condensation operation in \cite{SchumacherWestmoreland}. In general, the map
$|\psi\rangle\mapsto|\psi\rangle\circ|\varphi\rangle$
is not even an isometry, as the example after Definition~\ref{DefQConcat} shows.
Yet, maybe it is an isometry, and thus physically realizable, if it is restricted to prefix-free Hilbert spaces.

Another interesting question is if these spaces
are relevant for algorithmic information theory. This motivation is also mentioned in \cite{SchumacherWestmoreland}.
In particular, the role of these generalized prefix-free Hilbert spaces
should be clarified for the definition of prefix-free quantum computers and quantum algorithmic probability.

\section*{Acknowledgment}
One of us (M.M.) would like to thank Nihat Ay, Arleta Szko\l a, Tyll Kr\"uger,
Rainer Siegmund-Schultze, Christopher Witte and David Gross
for support and interesting discussions.

\vfill

\end{document}